\documentclass[%
prd,%
preprintnumbers,%
amsmath,%
amssymb,%
twocolumn]{revtex4}
\usepackage[utf8]{inputenc}

\usepackage{graphicx}
\usepackage{color}
\usepackage[dvipsnames]{xcolor}
\usepackage{epstopdf}
\usepackage{hyperref}
\usepackage{cancel} 

\newcommand{\bea}{\begin{eqnarray}}
\newcommand{\eea}{\end{eqnarray}}

\begin{document}

\title{\boldmath Enhancement of Direct $CP$ asymmetry in $Z'$ models}

\author{Shireen Niteen Gangal}
\email{shireen@prl.res.in}
\affiliation{Physical Research Laboratory, Ahmedabad, Gujrat, India.}

\begin{abstract}
We consider $CP$-violating $Z'$ models to account for the anomalies in $b \to s \ell \ell$ decays. Using the updated constraints from lepton flavor universality (LFU) violating ratios,  
$b \to s \mu \mu $ $CP$-conserving and $CP$-violating observables, $B_s-\bar{B}_s$ mixing and neutrino trident we obtain the favored parameter space of two classes of $Z'$ models generating the `1D' new physics scenarios with $C_9^{\rm NP} <0$ and $C_9^{\rm NP} = -C_{10}^{\rm NP}$. 
We show that the predictions of direct $CP$ asymmetry $A_{CP}$ in $B^+ \to K^+ \mu \, \mu$ decays close to the $c \bar{c}$ resonance region can be used to detect the presence of new $CP$ violating phases in the $Z'$ couplings.
The favored $1\,\sigma$ parameter space of $Z'$ models generating the scenario $C_9^{\rm NP} <0$ allows for an enhancement in
the integrated $A_{CP}$ in $q^2 =[8,9]\,\rm{GeV^2}$ bin up to $\pm 10\%$ and in the 
$q^2 =[6,7]\,\rm{GeV^2}$ bin up to $\pm 5\%$. 
We find that for a negative value of the phase of $J/\psi$ resonance, although such an enhancement is possible in $Z'$ models generating the scenario $C_9^{\rm NP} = -C_{10}^{\rm NP}$, the
favored parameter space prefers only positive values of $A_{CP}$. Hence, a future measurement of direct $CP$ asymmetry in these bins can potentially constrain the size of new $CP$ violating phases, and help
in distinguishing between the two classes of $Z'$ models.
\end{abstract}


\maketitle

\section{Introduction}
There have been intriguing discrepancies between the measurements of a few observables in $b \to s \ell \ell$ decays and their Standard Model predictions over the last few years.
Among these, the most significant ones are the multiple hints of violation of the lepton flavor universality (LFU), which is embedded in the gauge structure of the SM. 
The LFU has been tested through the measurements of the ratio observables $R_{K} = \Gamma (B^+ \to K^+ \mu^+ \mu^-)/\Gamma(B^+ \to K^+ e^+ e^-)$ and  $R_{K^*} = \Gamma (B^0 \to K^{0*} \mu^+ \mu^-)/\Gamma(B^0 \to K^{0*} e^+ e^-)$ which are lower than the SM prediction of $\sim 1$ by $3.1\,\sigma$ and $2.5\,\sigma$ respectively~\cite{LHCb:2021trn,Aaij:2017vbb}. Recent measurements of LFU ratios 
defined in the channels $B_d^0 \to K_S^0\,  \mu^+ \mu^-$ and $B^+ \to K^{*+}\mu^+ \mu^-$ show similar deficits, though not so significant, at $1.5\,\sigma$~\cite{LHCb:2021lvy}. 
Apart from the LFU ratios, the branching ratio of $B_s \to \phi \mu \mu$ measured by LHCb also exhibits a deficit compared to the SM prediction at the level of $3.5\,\sigma$~\cite{LHCb:2015wdu,LHCb:2021zwz}. 
Further, the updated measurement of the angular observable $P_5^\prime$ by LHCb, defined using the four-fold angular distribution of $B_d^0 \to K^{*0} (\to K^+ \pi-) \mu^+ \mu^-$ shows a disagreement
with the SM prediction at $3.7\,\sigma$~\cite{LHCb:2013ghj,LHCb:2015svh,LHCb:2020lmf,Descotes-Genon:2013vna}. 

These anomalies have been addressed in two ways, either using effective field theories (EFT) by including all possible new dimension-six operators,
or building specific new physics models. For the EFT analyses, global fits are performed in a model-independent way to all the $ b \to s \ell \ell$
data, in order to find the preferred Lorentz structure of the new physics (NP) operators. 
In the case of one-parameter scenarios, global fits show a preference to the Wilson Coefficient (WC) combinations, $C_9^{\rm NP} < 0$, or $C_9^{\rm NP} = -C_{10}^{\rm NP}$,
corresponding to the NP operators $O_9 = (\bar{s} \gamma_\mu P_L b) \, (\bar{\ell} \gamma^\mu  \ell) $ and $O_{10}= (\bar{s} \gamma_\mu P_L b) \, (\bar{\ell} \gamma^\mu \gamma^5 \ell) $ \cite{Descotes-Genon:2013wba,Altmannshofer:2013foa,
Alok:2017jgr,Alok:2019ufo,Altmannshofer:2021qrr,Carvunis:2021jga,Alguero:2021anc,Geng:2021nhg,Hurth:2021nsi,Angelescu:2021lln}.
Global fits with two non-zero real NP WCs also have significant pulls, however they continue to indicate a strong preference to the presence of $C_9^{\rm NP} < 0$.
In the context of models, the two classes proposed to account for these anomalies are $Z'$ models, and models with leptoquarks (LQ). 
In simplified $Z^\prime$ models, the $Z'$ boson couples to $\bar{ s}b$ and $\mu$ at tree-level and can contribute to both the favored `1D' scenarios $C_9^{\rm NP}<0$ and $C_9^{\rm NP} = -C_{10}^{\rm NP}$, while the LQ models can only generate the latter.

Although these NP scenarios with real WCs are the preferred ones, 
in general these NP WCs can also be complex, thereby giving rise to new sources of $CP$ violation. 
Moreover, since $CP$ violating effects in $ b \to s $ decays are suppressed in the SM, 
these are promising channels to look for new sources of $CP$ violation. The new $CP$ violating phases are very weakly constrained as there are only a few 
measurements of $CP$-violating observables. Global fits with complex NP WCs have been
performed in Refs.~\cite{Alok:2017jgr,Carvunis:2021jga, Altmannshofer:2021qrr,SinghChundawat:2022zdf}, and a few studies in the past obtained constraints on the parameter space
of $Z'$ and leptoquark models \cite{Alok:2017jgr,DiLuzio:2019jyq, Kosnik:2021wyp,Alok:2019xub}.
The most relevant $CP$ violating observables are mixing-induced
$CP$ asymmetry $A_{CP}^{\rm mix}(B_s \to J/\psi \phi)$ 
and $CP$ asymmetric observables in the $b \to s \ell \ell$ sector.
The latter include direct $CP$ violation in $B \to K^{(*)} \mu \mu$, and $CP$ asymmetric angular
observables $A_7$, $A_8$ and $A_9$ measured by LHCb \cite{LHCb:2014mit, LHCb:2015svh}, which still have large uncertainties and are consistent with zero. 

The measurement of $A_{CP}$ in $b \to s \ell \ell$ decays is difficult partly because it is very small, and there are asymmetries in production rate and detection efficiencies that
affect the measurements. The decay $B \to J/\psi K^*$ which has negligible direct $CP$ asymmetry is used as a control mode to reduce these asymmetries. The LHCb analysis in Ref.~\cite{LHCb:2014mit} 
is performed in the $0.1 \le q^2 \le 19.0\,\rm{GeV}^2 $ bin, and the regions near the $\phi(1020)$, $J/\psi$ and $\psi(2S)$ resonances are removed.
However, it has been recently shown that the measurement of $A_{CP}$ near the $J/\psi$ , $\psi(2S)$ resonances is more interesting as it could lead 
to a potential large enhancement in the presence of new $CP$ violating phases \cite{Becirevic:2020ssj}. 
Motivated by this, we study the imprints of a class of $CP$-violating non-universal $Z'$ models generating the NP scenarios 
$C_9^{\rm NP} <0$ and $C_9^{\rm NP} = -C_{10}^{\rm NP}$
on the direct $CP$ asymmetry in $B^+ \to K^+ \mu^+ \mu^-$. For this, we obtain the favored $1\sigma$ parameter space of $Z'$ models with
complex couplings, using the updated measurements from $b\to s \ell \ell$ observables, $B_s-\bar{B}_s$ mixing and
neutrino trident. 

This work is organized as follows: in the next section we consider the effective Hamiltonian for $b \to s \ell \ell$ decays
and define the NP WCs $C_9^{\rm NP}$ and $C_{10}^{\rm NP}$ in terms of the $Z'$ couplings,
through matching. In sec.~\ref{constraint}, we list the measurements that constrain the complex $Z'$ couplings and describe our fit methodology. In sec.~\ref{ACP}, we discuss
how we parametrize the region near charmonium resonance, and 
present the predictions of integrated $A_{CP}$ in the allowed parameter space of the $Z'$ model. We conclude in sec.~\ref{conclusions}. 

\section{The $Z'$ model and $\mathbf{b \to s \ell \ell}$ transitions}
\label{Z'}
The effective Hamiltonian for  $b\rightarrow s \mu \,\mu$ transitions is given by,
\begin{align} \nonumber
  \mathcal{H}_{\rm eff}^{\rm b \to s \ell \ell} &= -\frac{ 4 G_F}{\sqrt{2}} V_{ts}^* V_{tb}
  \bigg[ \sum_{i=1}^{6}C_i \mathcal{O}_i\\
  &  + C_7^{\rm SM}\frac{e}{16 \pi^2}[\overline{q} \sigma_{\mu \nu}
      (m_s P_L + m_b P_R)b] F^{\mu \nu}  + C_8 {\mathcal O}_8 \nonumber\\
     & + C_9 \frac{\alpha_{\rm em}}{4 \pi}
    (\overline{s} \gamma^{\mu} P_L b)(\overline{\mu} \gamma_{\mu} \mu)  \nonumber\\    
  &  + C_{10} \frac{\alpha_{\rm em}}{4 \pi}
    (\overline{s} \gamma^{\mu} P_L b)(\overline{\mu} \gamma_{\mu} \gamma_{5} \mu) 
    \bigg] + \text{h.c.} \;,
\label{eq:Heff}
\end{align}
where $G_F$ is the Fermi constant and $V_{ij}$ are the Cabibbo-Kobayashi-Maskawa
(CKM) matrix elements. The contribution from the operators $O_{i=1...6}$ are
included through the modification, $C_{7,8, 9}^{\rm SM} \to C_{7,8,9}^{\rm eff, SM}$.
The presence of new physics modifies the WCs corresponding to semi-leptonic
operators $O_9$ and $O_{10}$ as follows, $C_9 = C_9^{\rm eff, SM } + C_9^{\rm NP}$ and 
$C_{10} = C_{10}^{\rm eff, SM } + C_{10}^{\rm NP}$.  

The NP contribution to $B_s -\bar{B}_s$ mixing can be parameterized by the effective Hamiltonian,
\begin{align}
\mathcal{H}_{\rm eff}^{\Delta B=2} = -\frac{4G_F}{\sqrt{2}} V_{tb}V_{ts}^* ( C_1^{bs} (\bar{s}\gamma_\mu b)^2 ) + \text{h.c.} 
\end{align}
where $C_1^{bs}$ is modified as, $C_1^{bs} = C_1^{bs, \rm SM} + C_1^{ bs, \rm NP}$.

Recent global fits with one NP parameter indicate that the most preferred scenarios are:
Scenario I: $C_9^{\rm NP} < 0$ and Scenario II:
$C_9^{\rm NP} = -C_{10}^{\rm NP}$, assuming that NP affects only the muon sector. 
These two favored NP scenarios can be generated in $Z'$ models. We consider a model with a $Z'$ boson associated with $U(1)^\prime$ extension of the SM. The couplings of such a $Z'$ boson relevant
for $b \to s \ell \ell$ decays are given by, 
\begin{align}
\mathcal{L}_{Z'} &\supset  Z_\alpha J_{Z'}^\alpha  \nonumber\\
   &= (g^{\mu\mu}_L\, \bar{L}_2 \gamma^{\alpha}P_L L_2 + g^{\mu \mu}_R\, \bar{e}_2 \gamma^{\alpha}P_R\, e_2  + g^{bs}_L\, \bar{s} \gamma^{\alpha}P_L b ) Z'_\alpha \,,
\label{eq:Jalpha}
\end{align}
where  $L_2  (e_2)$ are the second generation lepton doublets (singlets) and $g_L^{\mu \mu} (g_R^{\mu \mu})$ are left-handed 
(right-handed) couplings of $Z'$ to muons, and $g_L^{bs}$ are couplings to quarks. We only consider left-handed couplings in the quark sector. 
Since the $Z'$ is much heavier, it can be integrated out to get an effective Hamiltonian
with relevant four fermion interactions given by,
\begin{align}
  \mathcal{H}_{\rm eff}^{Z'} &= \frac{1}{2M^2_{Z'}}J_{\alpha}J^{\alpha}  \nonumber\\
&=  \frac{g^{bs}_L}{M^2_{Z'}} \left(\bar{s}\gamma^{\alpha}P_L b\right)
  \left[\bar{\mu}\gamma_{\alpha}\left(g^{\mu\mu}_L P_L 
   + g^{\mu\mu}_R P_R\right)\mu \right]  \nonumber \\
&   +  \frac{\left(g^{bs}_L\right)^2}{2M^2_{Z'}}\left(\bar{s}\gamma^{\alpha}P_L b\right)\left(\bar{s}\gamma_{\alpha}P_L b\right)
 \nonumber\\
& + \frac{g^{\mu\mu}_L}{M^2_{Z'}} \left(\bar{\nu}_{\mu}\gamma_{\alpha}P_L\nu_{\mu}\right) \left[\bar{\mu}\gamma^{\alpha}\left(g^{\mu\mu}_L P_L + g^{\mu\mu}_R P_R\right)\mu\right]\,.
 \label{eq:HeffZp}
\end{align}
Here, the first term contributes to $ b \to s \ell \ell$ transitions, the second term to $B_s-\bar{B}_s$ mixing and the last term to neutrino trident production $\nu_\mu N \to \nu_\mu N \mu^+ \mu^- $.
The NP WCs in the $Z'$ model obtained by matching eq.~\ref{eq:HeffZp} on to the $b \to s \ell \ell$ effective Hamiltonian eq.~\ref{eq:Heff} are,
\begin{eqnarray}
  C^{\rm NP}_9 &=& -\frac{\pi}{\sqrt{2}G_F\alpha V_{tb}V^*_{ts}} \frac{g_L^{bs}(g_L^{\mu\mu}+g_R^{\mu\mu})}{M^2_{Z'}}\,,\\
  C^{\rm NP}_{10} &=& \frac{\pi}{\sqrt{2}G_F\alpha V_{tb}V^*_{ts}} \frac{g_L^{bs}(g_L^{\mu\mu}- g_R^{\mu\mu})}{M^2_{Z'}}\,,\\
C_1^{bs} &=& \frac{1}{4 \sqrt{2} G_F M_{Z'}^2}\Big( \frac{g_L^{bs}}{V_{tb} V_{ts}^*}\Big)^2\,.
  \label{bqllNP}
\end{eqnarray}
The $Z'$ models that can generate Scenario I require $g_L^{\mu \mu} = g_R^{\mu \mu}$, and those giving rise to Scenario II require $g_R^{\mu \mu} = 0$. The scenario I, with $C_9^{\rm NP}$ alone, can be generated in $Z'$ models with an additional
$U(1)_{L_\mu - L_\tau}$ symmetry as shown in Ref.~\cite{Altmannshofer:2014cfa, Crivellin:2015mga}. The scenario II, $C_9^{\rm NP} = -C_{10}^{\rm NP}$, can be generated
in models which assume couplings only to left-handed leptons \cite{AristizabalSierra:2015vqb}.

\section{Constraints and Fit Methodology} 
\label{constraint}

\begin{figure*}[t]
\includegraphics[width=0.45\textwidth]{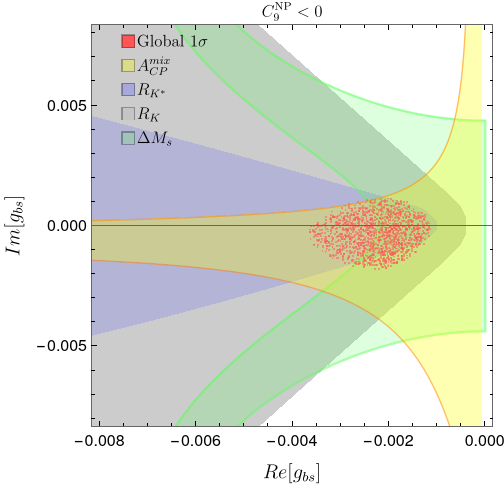}
\hspace{0.02\textwidth}
\includegraphics[width=0.45\textwidth]{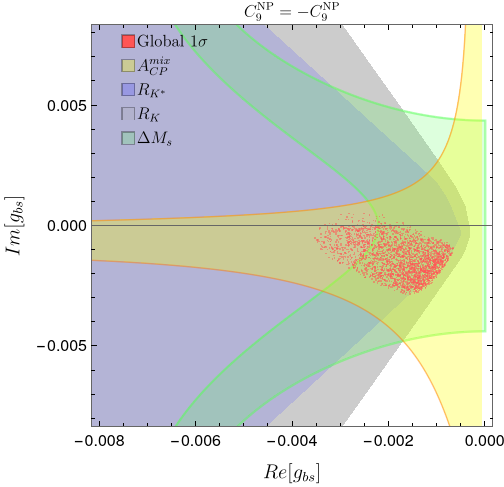}
\caption{The $1\sigma$ allowed region (red) in the parameter space of complex $g_{bs}$ coupling for $Z'$ models generating `1D' scenarios with $C_9^{\rm NP} < 0$ (left) and $C_9^{\rm NP} = -C_{10}^{\rm NP} $ (right). The $1\,\sigma$ constraints from 
$B_s$ mixing observable $\Delta M_s$ (green), mixing induced $CP$ asymmetry $A_{CP}^{\rm mix}$ (yellow) and LFU ratios $R_{K^*}$ (blue) and $R_K$ (gray) are also shown.}
\label{fig:Zprime}
\end{figure*}

The observables in $b \to s \mu^+ \mu^-$ sector constrain the product of the couplings $g_{L}^{bs}g^{\mu\mu}_{L,R}$. 
We follow the fit methodology adopted in Ref.~\cite{Alok:2019ufo} and use the updated observables listed in Ref.~\cite{Alok:2022pjb}. Since the $Z'$ couplings are complex we include in the fit additional constraints
from $CP$ asymmetric angular observables measured by LHCb \cite{LHCb:2015svh}. The $\chi^2$ function for the $b \to s \mu^+ \mu^-$ observables is given by,
\begin{equation}
\chi^2_{b \to s \mu \mu}(C_i) = [\mathcal{O}_{\rm th}(C_i) -\mathcal{O}_{\rm exp}]^T  \mathcal{C}^{-1} 
[\mathcal{O}_{\rm th}(C_i) -\mathcal{O}_{\rm exp}],
\label{chi-2d}
\end{equation} 
where  $C_i = C^{\rm NP}_{9,10}$. 
The theoretical predictions of $b \to s \mu^+ \mu^-$ observables calculated using {\tt flavio} \cite{Straub:2018kue} are denoted by $\mathcal{O}_{\rm th}(C_i)$ and the corresponding experimental measurements
by $\mathcal{O}_{\rm exp}$.  
The total covariance matrix $\mathcal{C}$ is obtained by adding the individual theoretical and experimental covariance matrices.

The $Z'$ quark coupling $g_{bs}$ receives strong constraints from the measurement of $B_s - \bar{B}_s$ mixing through the expression \cite{DiLuzio:2019jyq},
\begin{equation}
\frac{\Delta M_s ^{\rm SM + NP}}{\Delta M_s^{\rm SM}} = \Big| 1 + \frac{\eta^{6/23}}{R_{\rm loop}^{SM}} C^{bs}_1  \Big|
\end{equation}
where $\eta = \alpha_s(\mu_{NP})/\alpha_s(m_b)$ and 
\begin{equation}
R^{SM}_{\rm loop} = \frac{\sqrt{2}G_F M_W^2 \eta_B S_0(x_t)}{16 \pi^2} = 1.31 \times 10^{-3}
\end{equation}
The measurement of mixing-induced $CP$ asymmetry in $B_s \to J/\psi \phi$ can be used to constrain $Im[g_{bs}]$, which in the presence of NP is given by,
\begin{align}
A_{CP}^{\rm mix} = \sin{(\phi_\delta - 2 \beta_s)}
\end{align}
where,
\begin{align}
\phi_\delta = \arg\Big(1 + \frac{C^{bs}_{1}}{R^{SM}_{\rm loop}} \Big)\,.
\end{align}

\begin{figure*}[t]
\includegraphics[width=0.46\textwidth]{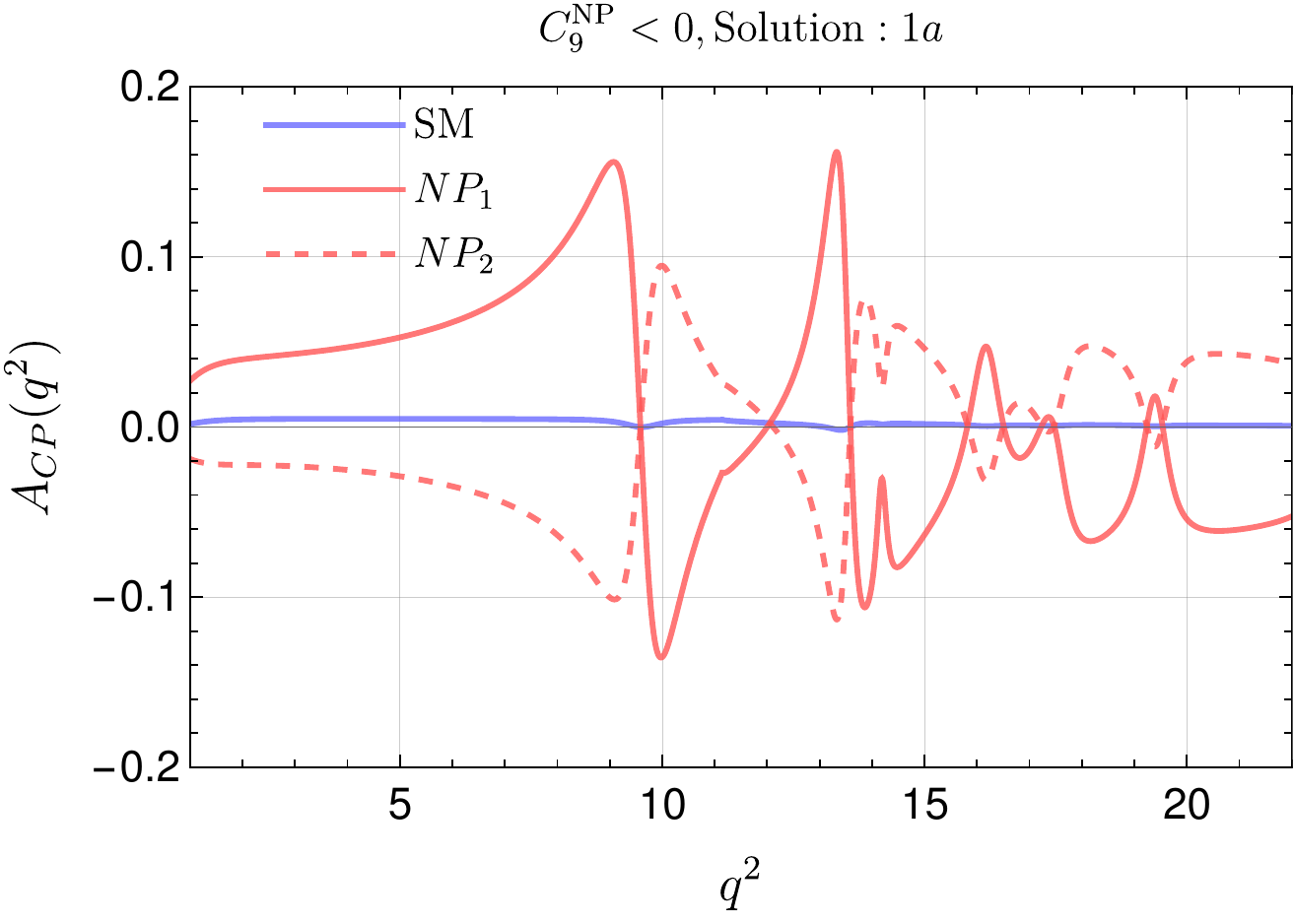}
\hspace{0.02\textwidth}
\includegraphics[width=0.46\textwidth]{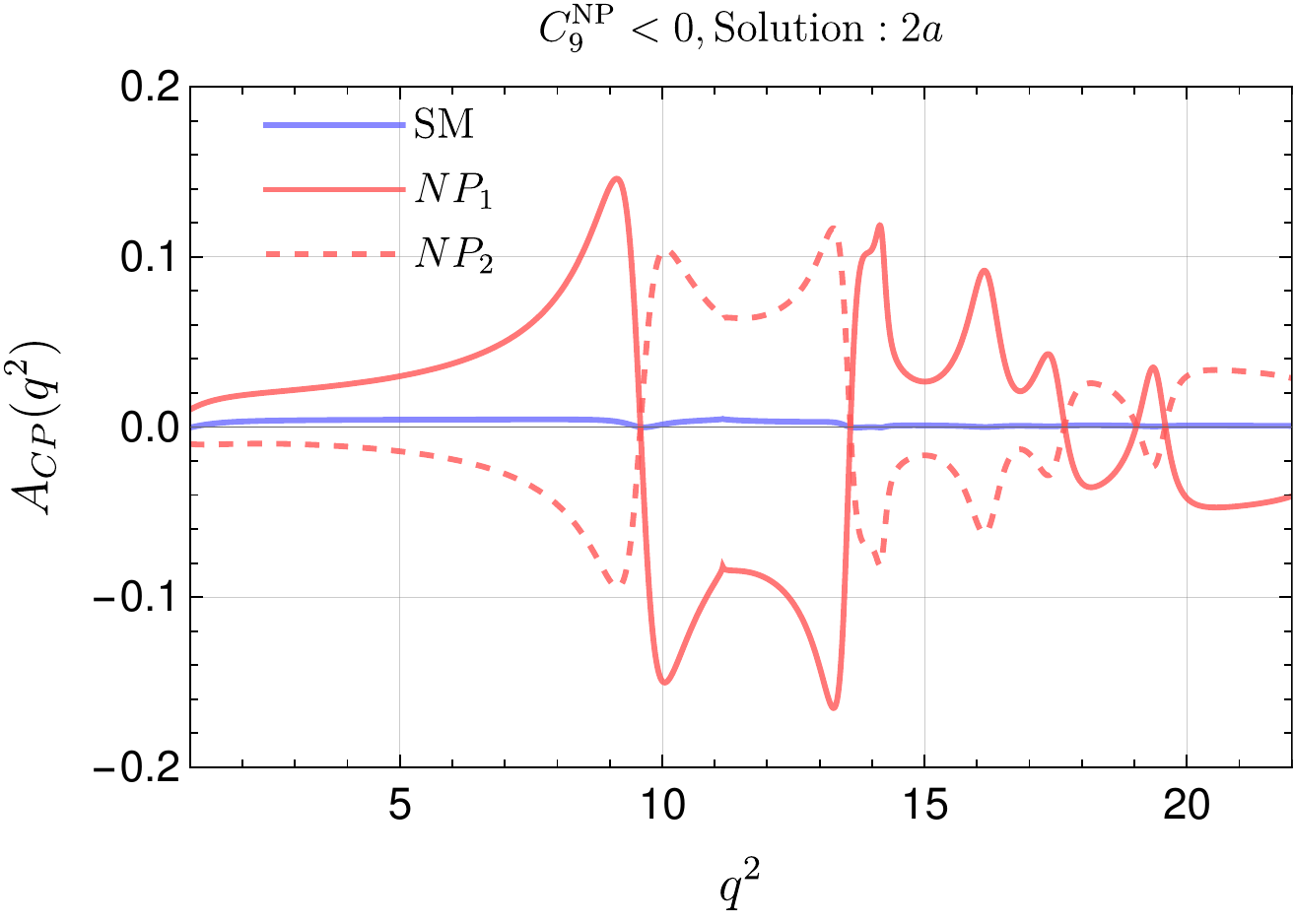}
\caption{Predictions of $A_{CP}(q^2)$ in the full $q^2$ region for the benchmark points NP1: $C_9^{\rm NP}= -0.93 -0.75\,i$ and NP2: $C_9^{\rm NP} = -0.89 + 0.50\, i$ corresponding to the $1\sigma$ maximal allowed $Z'$ couplings
for Scenario I. 
The left and right panels are for the two strong phase choices in solution 1a and 
solution 2a respectively.}
\label{fig:ACPI}
\end{figure*}
The lepton couplings $g_{L,R}^{\mu \mu}$ can also be constrained using neutrino trident production $\nu_\mu N \to \nu_\mu N \mu^+ \mu^-$, since by $SU(2)_L$ invariance the $Z'$ couples with neutrinos
via the same coupling as muons. The cross section for neutrino trident can be parameterized as, 
\begin{eqnarray}
R_\nu =  \frac{\sigma}{\sigma_{\rm SM}} &=& \frac{1}{1+(1+4s^2_W)^2}\Bigg[\left(1+ \frac{v^2g^{\mu\mu}_L(g^{\mu\mu}_L-g^{\mu\mu}_R)}{M^2_{Z'}}\right)^2 \nonumber\\
&&
+ \left(1+4s^2_W+\frac{v^2g^{\mu\mu}_L(g^{\mu\mu}_L+g^{\mu\mu}_R)}{M^2_{Z'}}\right)^2\Bigg],
\label{trident}
\end{eqnarray}
where $v=246$ GeV, $s_W = \sin\,\theta_W$ and we use $R_\nu^{\rm exp} = 0.82 \pm 0.28$. 

Combining the constraints from the above measurements, we find the best-fit values for the $Z'$ couplings, assuming $M_{Z'} = 1\, \mathrm{TeV}$, to be:\\
(Re$[g_L^{bs}]$, Im$[g_L^{bs}], g_L^{\mu \mu})$ = $(-2.6 \times 10^{-3}, -0.4 \times 10^{-3}, 0.28)$,
(Re$[g_L^{bs}]$, Im$[g_L^{bs}], g_L^{\mu \mu}$) 
= $(-1.5 \times 10^{-3}, -1.8\times 10^{-3},0.47)$\\ 
for scenario I and II respectively.

Fig.~\ref{fig:Zprime} depicts the $1\sigma$ favored region in the parameter space of complex quark coupling $g_{bs}$
for $C_9^{\rm NP} < 0$  (left) and  $C_9^{\rm NP} = -C_{10}^{\rm NP}$ scenarios (right). 
The NP $Z'$ couplings improve the fit for the $ b \to s \mu \mu$ observables, with 
$\Delta \chi^2 = \chi^2_{\rm SM}- \chi^2_{\rm NP} \sim 42$. 
Marginalizing over $g_L^{\mu \mu}$, we find that the favored 1$\sigma$ region in the plane of ${\rm Re}[g_{bs}]- {\rm Im}[g_{bs}]$ is consistent with the $1\sigma$ bounds from $R_K$, $A_{CP}^{\rm mix}$ and $\Delta M_s$. 
The strongest bound on the imaginary part of these $Z'$ quark couplings come from the measurements of $\Delta A_{CP}^{\rm mix}$, 
except at small negative values of ${\rm Re}[g_{bs}]$ , where the
constraint from $R_{K^{*}}$ is the dominant one. We find that imaginary couplings as large as the real ones are still allowed by the current data.  
We use the ratio $\Delta M_s^{\rm SM}/ \Delta M_s^{\rm exp} = 1.04 \pm 0.07 $ \cite{DiLuzio:2019jyq}, which corresponds to the current experimental value of $\Delta M_s^{\rm exp} = 17.757 \pm 0.021$, 
and the SM prediction given by the weighted average of sum rule and FLAG 2019 prediction, $\Delta M_s^{\rm SM} = (18.4 \pm 1.2 ) \,\rm{ps}^{-1}$. The 
preference to Im[$C_9^{\rm NP}]$ is more pronounced in the case of FLAG 2019 SM prediction, $\Delta M_s^{\rm SM} = (20.1 \pm 1.6 ) \rm{ps}^{-1}$ which has a much higher central value than the weighted average.
The bounds from $CP$ asymmetric angular observables $A_7, A_8, A_9$ show some preference
towards Im[$C_9^{\rm NP}] < 0$, and this is more prominent in the case of $C_9^{\rm NP} = -C_{10}^{\rm NP}$ scenario which shifts the best-fit towards larger negative Im$[g_{bs}]$ values, as can be seen in fig.~\ref{fig:Zprime} .

\section{$A_{CP}$ Predictions}
\label{ACP}
\begin{figure*}[t]
\includegraphics[width=0.42\textwidth]{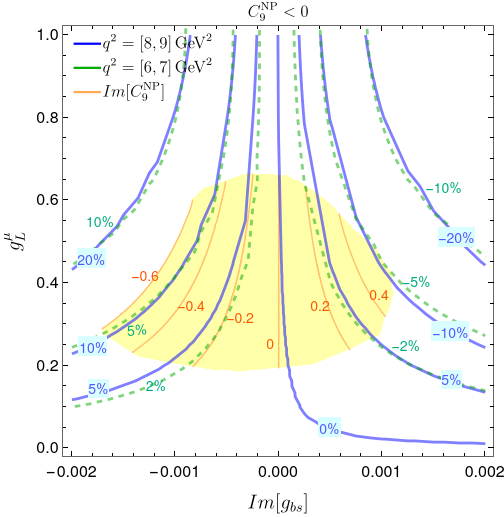}
\hspace{0.05\textwidth}
\includegraphics[width=0.42\textwidth]{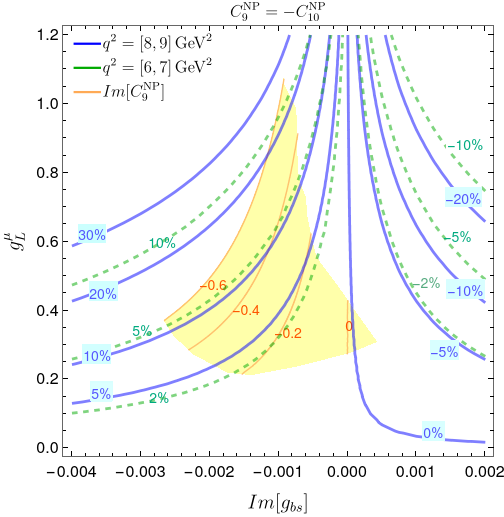}
\caption{The $1\sigma$ favored region shown in yellow on the parameter space of $Z'$ couplings $Im[g_{bs}]$ and $g_L^{\mu}$, along with the corresponding $Im[C_9^{\rm NP}]$ values shown by contours in orange for Scenario I (left) and 
II (right). The green dashed 
and blue contours denote the integrated $A_{CP}$ values in the $q^2 = [6,7]\,\mathrm{GeV^2}$ and $q^2 = [8,9]\,\mathrm{GeV^2}$ bins respectively.}
\label{fig:ACPCont}
\end{figure*}

We now study the implications of complex $Z'$ couplings on the predictions of direct $CP$ asymmetry in $B^+ \to K^+ \mu^+ \mu^-$ decays. Since the SM
prediction of $A_{CP}$ is very small $O(10^{-3})$, a non-zero measurement of $A_{CP}$ could indicate presence of new physics. The direct $CP$ asymmetry is defined as,
\begin{align}
A_{CP}(q^2) = \frac{d\bar{\Gamma}(B^- \to K^- \mu \mu)/dq^2 - d\Gamma (B^+ \to K^+ \mu \mu)/dq^2}{ d\bar{\Gamma}(B^- \to K^- \mu \mu)/dq^2 + d\Gamma (B^+ \to K^+ \mu \mu)/dq^2}
\end{align}
where the differential decay rate for $B^+ \to K^+ \mu \mu$ and the corresponding form factors are taken from Ref.~\cite{Bailey:2015dka}.

A non-zero
$A_{CP}$ requires an intereference between two amplitudes with different strong and weak phases. This is possible due to an interefernce between
the phase of the $Z'$ couplings and the strong phases in the $c \bar{c}$ resonance region. 
The effect of the presence of $c \bar{c}$ resonances enter
via the process $B \to V_{c\bar c} \to \ell \ell$, where $V_{c\bar{c}}$ can be any of $J/\psi, \psi(2S), \psi(3770), \psi(4040), \psi(4160)$ or $\psi(4415)$.
These long-distance effects can be modelled theoretically via a sum over Breit-Wigner (BW)
poles as follows, \cite{Kruger:1996cv}
\begin{equation}
\label{eq:Vcc_shift}
C_9 ~\to~ C_9 ~-~ \frac{9 \pi}{\alpha^2} \, \bar C \,  \sum_V 
 |\eta_V| e^{i \delta_V} \frac{\hat{m}_V \, \mathcal B(V \to \mu^+ \mu^-) \, \hat{\Gamma}_{\textrm{tot}}^V}{\hat{q}^2 - \hat{m}_V^2 + i \hat{m}_V \hat{\Gamma}^V_{\textrm{tot}}}~.
\end{equation}
where $\bar C = \bar{C}_1 + \bar{C}_{2}/3 + \bar{C}_3 + \bar{C}_{4}/3 + \bar{C}_5 + \bar{C}_{6}/3$ and $|\eta_V| =2.5$.  The masses, branching ratios and decay widths of the resonances are taken from Ref.~\cite{ParticleDataGroup:2020ssz}. 
We use $C_9^{\rm SM} = C_9^{\rm eff} = C_9 + Y(q^2)$ where $Y(q^2)$ is taken from Refs.~\cite{Beneke:2020fot, Beneke:2001at}. The values of the strong phases $\delta_j$ are taken from LHCb analysis \cite{LHCb:2016due},
wherein these phases were determined through a fit to the full dimuon mass spectrum, using a model for the resonances in the form of a Breit-Wigner function.
The fit leads to four possible combinations such that the sign of the phase of $J/\psi$ is negative (Branch A) or positive (Branch B), and the $\psi(2S)$ phase can have either sign. 
The best fits for these are,
\begin{align} 
&\text{Branch A}: \nonumber \\
&\text{Solution 1a}:  \delta_{J/\psi} = -1.66 \,\,\,,\,\, \delta_{\psi(2S)} = -1.93 \nonumber \\ 
&\text{Solution 2a}:  \delta_{J/\psi} = -1.50 \,\,,\,\, \delta_{\psi(2S)} = 2.08 \,.\\
&\text{Branch B}:\nonumber \\ 
&\text{Solution 1b}:  \delta_{J/\psi} = 1.47 \,\,\,,\,\, \delta_{\psi(2S)} = -2.21  \nonumber \\ 
&\text{Solution 2b}:  \delta_{J/\psi} = 1.63 \,\,,\,\, \delta_{\psi(2S)} = 1.80 \,.
 \label{eq:phase}
\end{align}
It is sufficient to consider one of the branches, as the other branch only flips the signs of $A_{CP}$ in the presence of new complex phases. 
This is because the phase of $J/\psi$ changes sign in Branch B, and $A_{CP}$ prediction depends on the NP complex phase and the strong phase
in the resonance region as, $A_{CP} \propto {\rm Im}[C_9^{\rm NP}] \sin{\delta_V}$ \cite{Becirevic:2020ssj}.
We consider Branch A and solution 1a in our analyses.

In fig.~\ref{fig:ACPI}, we show the predictions of $A_{CP}(q^2)$ in the full $q^2$ region for scenario I with $\rm{Re}[C_9^{\rm NP}] < 0$ generated in $Z'$ model, for two different phase choices. To get an estimate of the maximum deviation
in $A_{CP}(q^2)$ we consider the following two benchmark points 
corresponding to the maximum allowed values of ${\rm Im}[g_{bs}]$ within $1\sigma$:\\
NP1: ($g_L^{bs}, g_L^{\mu \mu}) = \{(-2.0 + 1.1\, i) \times 10^{-3}, 0.37 \}$ \\
NP2: ($g_L^{bs}, g_L^{\mu \mu}) = \{(-2.2 - 1.8\, i)\times 10^{-3} ,  0.35 \}$ \\
These benchmark points correspond to $C_9^{\rm NP}$ values $-0.93 - 0.75\, i$ and $-0.89 + 0.50\, i$ respectively.
It can be seen from the plot that the values of $A_{CP}(q^2)$ can be as large as $\sim 15\%$ very close to the resonance peaks and an enhancement of $\sim 5- 10\%$ 
seems possible in the region $q^2 = [6-8] \mathrm{GeV}^2$.
If future measurements prefer solution 1a with same signs for the phases of $J/\psi$ and $\psi(2S)$, then a measurement of $A_{CP}$ in the $q^2 = [6,10]\,\mathrm{GeV}^2$ bin could tell apart NP1 and NP2 due
to the flip in the sign of $A_{CP}$. 
On the other hand, if solution 2a is preferred, then the measurement of $A_{CP}$ in the $q^2 = [14,20]\,\mathrm{GeV}^2$ bin can also distinguish between NP1 and NP2, which 
gives rise to positive and negative $A_{CP}$ values respectively.


LHCb has measured $A_{CP}$ in 17 bins in the $q^2=  [ 0.1 -22] \mathrm{GeV^2} $ region, while vetoeing the regions $[8,11]$ and [12.5,15.0] around the $c \bar{c}$ resonances. 
While there is a larger enhancement in the $A_{CP}(q^2)$ predictions near the $c\bar{c}$ resonances as shown in fig.~\ref{fig:ACPI}, this also extends further away from the resonance peaks up to $q^2 = 6\, \mathrm{GeV}^2$. 
We obtain NP predictions in the region $q^2 = [8,9]\, \mathrm{GeV^2}$ near the $c\bar{c}$ resonance, and also in $q^2 = [6,7]\, \mathrm{GeV^2}$ bins where LHCb measurement already exists,
albeit with larger uncertainties. 
The binned $CP$ asymmetry is defined as,
\begin{align}
A_{CP}[q^2_{\rm min}, q^2_{\rm max}] = \frac {\bar{\Gamma}(B^- \to K^- \mu \mu) - \Gamma (B^+ \to K^+ \mu \mu)}{ \bar{\Gamma}(B^- \to K^- \mu \mu) + \Gamma (B^+ \to K^+ \mu \mu)}
\end{align}
where $\Gamma = \int_ {q^2_{\rm min}}^{q^2_{\rm max}} d\Gamma/dq^2$ is the binned decay rate. 

In Fig.~\ref{fig:ACPCont}, we show the predictions of integrated $A_{CP}$ superposed on the $1\,\sigma$ allowed region in the plane of $Z'$ couplings ${\rm Im}[g_{bs}]$ and $g_L^{\mu \mu}$ for NP scenarios I and II. We obtain the NP predictions of $A_{CP}$
in the $q^2 = [6,7]\,\mathrm{GeV^2}$ and $q^2 = [8,9]\,\mathrm{GeV^2}$ bins by varying the values of couplings ${\rm Re}[g_L^{bs}], {\rm Im}[g_L^{bs}]$ and $g_L^{\mu \mu}$ in their $1\sigma$ allowed regions for scenario I and II. We find that
the $1\,\sigma$ favored region in scenario I allows for an enhancement in $A_{CP}$ both in positive and negative directions up to $10\%$ in the $q^2 = [8,9]\,\mathrm{GeV^2}$ bin, while only positive values are allowed for scenario II. 
The negative and positive values of $A_{CP}$ arise from the positive and negative NP phases, respectively. This is true for both the $J/\psi$ and $\psi(2S)$ phase choices in Solution 1a and Solution 2a (eq.~\ref{eq:phase}),
since the sign flip of $A_{CP}$ in Solution 2a happens only above $q^2 = 11 \,\mathrm{GeV}^2$. The favored region also allows for $A_{CP}$ enhancement of up to $\pm 5\%$ in the $q^2 = [6,7]\,\mathrm{GeV^2}$ bin
for scenario I, and $+5\%$ in scenario II. For the $q^2 = [10,11]\,\mathrm{GeV^2}$ bin, we find an enhancement of the same order as in $q^2 = [8,9]\,\mathrm{GeV^2}$ bin but with the $A_{CP}$ signs flipped for both the phase choices in branch A.
There is still an ambiguity in the sign of $A_{CP}$ as this sign would flip depending on whether the phase of $J/\psi$ is negative 
(branch A) or positive (branch B).
However, our choice of branch A is also motivated from the
theory prediction of Ref.~\cite{Khodjamirian:2012rm}, wherein
the non-local hadronic amplitude for $ B \to K \ell \ell$ is split into contributions from different flavors, and expressed in terms of dispersion relations. 
Varying the phases of $J/\psi$, $\psi(2S)$, 
a fit to the complex parameters in the dispersion relations is repeated, and a best-fit for these parameters is obtained for a negative value of the $J/\psi$ phase. 
With this choice, we find that any non-zero negative $A_{CP}$ values up to a few percent both in the $q^2 = [6,7]\,\mathrm{GeV^2}$ and 
$q^2 = [8,9]\,\mathrm{GeV^2}$ bins would indicate a stronger preference to $Z'$ models generating scenario I alone. 
With this choice of the phase of $J/\psi$ resonance, positive values of $A_{CP}$ point towards ${\rm Im}[C_9^{\rm NP}] <0$, which
is also preferred by the current measurements of $CP$ asymmetric angular observables $A_7$, $A_8$ and $A_9$.

\section{Conclusions}
\label{conclusions}

The accumulating discrepancies between the SM predictions and experimental measurements in $ b \to s \ell \ell$ decays indicate a strong 
preference to the new physics scenarios, $C_9^{\rm NP} <0$ and $C_9^{\rm NP} = -C_{10}^{\rm NP}$. 
The current $b \to s \ell \ell$ data allows for these WCs to be complex, and their imaginary parts to be atleast as large as the real ones, with some preference for
$Im[C_9^{\rm NP}] < 0$ arising from the measurements of $CP$ asymmetric angular observables in $B \to K^{0*} \mu \mu$ decays. 
Hence, the $Z'$ and leptoquark models generating these favored NP scenarios can have complex couplings, providing new sources of 
$CP$ violation. 

In this work, we determine the $1\sigma$ allowed region for the two classes of $Z'$ models generating `1D' scenarios $C_9^{\rm NP} < 0$ and $C_9^{\rm NP} = -C_{10}^{\rm NP}$ using constraints from the updated measurements of
all $ b \to s \ell \ell$ observables, $B_s-\bar{B}_s$ mixing, mixing-induced $CP$ asymmetry, and $CP$ asymmetric angular observables. 
We explore the possibility of using the predictions of direct $CP$ asymmetry near the $c\bar{c}$ resonance to distinguish between these two classes of $Z'$ models. We find that an enhancement in $A_{CP}$ up to $\pm 10\%$ and $\pm 5\%$ in the $q^2 = [8,9] \mathrm{GeV^2}$ and
$q^2 = [6,7] \mathrm{GeV^2}$ bins respectively
is allowed by the favored parameter
space of $Z'$ model generating the scenario $C_9^{\rm NP} <0$. The $1\sigma$ favored parameter space of these models for $C_9^{\rm NP} = -C_{10}^{\rm NP}$ scenario allows for only positive values of $A_{CP}$ in these bins, which is a potentially distinguishing feature. The sign of $A_{CP}$ flips depending on the choices
of the sign of $J/\psi$ phase as measured by LHCb, however our choice is also consistent with the sign of $J/\psi$ phase obtained in \cite{Khodjamirian:2012rm}.
Interestingly, the favored parameter space of leptoquark models lead to larger $(\sim 20\%)$ and only positive shifts in the values of $A_{CP}$ compared to $Z'$ models, hence a future
more precise measurement of $A_{CP}$ near the $c\bar{c}$ resonances can help in distinguishing between these
models favored by the current $B$ anomalies.

\begin{acknowledgments}
I would like to thank Namit Mahajan and Amol Dighe for useful discussions and comments on the manuscript.
\end{acknowledgments}

\bibliography{bibliography}

\end{document}